\documentclass[12pt]{article} 
\usepackage[xdvi]{graphicx} 
 
\begin{document}   
\newcommand{\todo}[1]{{\em \small {#1}}\marginpar{$\Longleftarrow$}}   
\newcommand{\labell}[1]{\label{#1}\qquad_{#1}} 

\rightline{DCPT-02/55}   
\rightline{hep-th/0207183}   
\vskip 1cm


\begin{center} 
{\Large \bf Stability of the non-extremal enhan\c con solution I:
perturbation equations}
\end{center} 
\vskip 1cm   
  
\renewcommand{\thefootnote}{\fnsymbol{footnote}}   
\centerline{\bf   
Apostolos Dimitriadis\footnote{Apostolos.Dimitriadis@durham.ac.uk} and Simon 
F. Ross\footnote{S.F.Ross@durham.ac.uk}}    
\vskip .5cm   
\centerline{ \it Centre for Particle Theory, Department of  
Mathematical Sciences}   
\centerline{\it University of Durham, South Road, Durham DH1 3LE, U.K.}   
  
\setcounter{footnote}{0}   
\renewcommand{\thefootnote}{\arabic{footnote}}


\begin{abstract}   
We consider the stability of the two branches of non-extremal enhan\c
con solutions. We argue that one would expect a transition between the
two branches at some value of the non-extremality, which should
manifest itself in some instability. We study small perturbations of
these solutions, constructing a sufficiently general ansatz for
linearised perturbations of the non-extremal solutions, and show that
the linearised equations are consistent. We show that the simplest
kind of perturbation does not lead to any instability. We reduce the
problem of studying the more general spherically symmetric perturbation to
solving a set of three coupled second-order differential equations.
\end{abstract}

\section{Introduction}     

A key issue in string theory is the r\^ole and physical interpretation
of singularities in supergravity solutions. Some singular solutions,
such as negative mass Schwarzschild, are genuinely
unphysical~\cite{Horowitz:sing}, and are simply excluded from
consideration; no corresponding source exists. String theory provides
resolutions of many other singularities through various
mechanisms. Recently, new singularity resolution mechanisms have played an important
part in the understanding of field theories with partially broken
supersymmetry in the AdS/CFT
correspondence~\cite{pol:sr,pilch:sr,kleb:sr,juan:sr,buchel:enh,evans:enh}.
A simple example of this new class of mechanisms is the resolution of
the repulson singularity of~\cite{Behrndt:enh,Kallosh:enh} by the
enhan\c con mechanism~\cite{JPP:enh}.

Generally, one of the simplest questions to consider from the bulk
spacetime side of the AdS/CFT correspondence is the finite-temperature
behaviour of the theory. One would expect that the theories with
reduced supersymmetry should have interesting phase structures.
At high temperatures, one would
expect to find that the partition function is dominated by a black
hole solution, and there may be some symmetry-breaking
phase transitions as the temperature decreases. Attempts to
investigate these issues by studying black hole solutions on the AdS
side were made
in~\cite{freed:bh,buchel:bh1,buchel:bh2,gubser:bh1,herzog:bh,gubser:bh2}.
Considerable progress was made on obtaining suitable black hole
solutions. However, because of the complexity of the setup, no exact
closed-form solutions are available.

In this paper, we will begin an investigation of the phase structure
associated with non-extremal black hole versions of the enhan\c con
solution of~\cite{JPP:enh}, using the simple explicit solutions
generalising the enhan\c con found in~\cite{JPP:enh,JMPR:enh}. We will
focus on studying whether the solutions have classical instabilities
which could provide the mechanism for transitions between them. We
analyse the linearised perturbation equations around the non-extremal
enhan\c con background, generalising the analysis of~\cite{Maeda:stab}
in the extreme case. Although the enhan\c con is somewhat different
from the asymptotically AdS cases, the underlying physics should be
similar. It would be interesting to extend our work to consider the
stability of the non-extremal fractional brane solutions
of~\cite{obers}, which are more closely related to asymptotically AdS
cases. 

In section~\ref{sec:solns}, we review the extremal and non-extremal
enhan\c con solutions. There are two branches of non-extremal
solutions, arising from an ambiguity of a choice of sign in the
solution of the supergravity equations. One branch joins on to the
extremal enhan\c con solution studied previously, and always has a
shell of branes outside the horizon. (The proportion of the energy
carried by the shell and by the black hole inside the shell in this
solution was not determined at this supergravity level; a better
understanding of the internal dynamics of the shell is required to
obtain 
a unique solution for given asymptotic charges.)
The other branch appears at a finite value of the non-extremality
parameter. Above this critical value of the non-extremality, both
types of solution are possible.  At large energies, the effects of the
D-brane charges should be negligible, so the solution with a horizon,
which for large mass is approximately the usual uncharged black hole
solution, has the right physical behaviour. On the other hand, if we
begin slowly adding energy to the extremal enhan\c con, we will obtain
a solution on the branch with a shell. We would expect that there is
some non-trivial transition between these two families of solutions as
we vary the energy.\footnote{Since the
enhan\c con is like a monopole solution, we expect the physics to be
similar to that of the Einstein-Yang Mills Higgs system
(see~\cite{weinberg} and references therein). For any given value of
the asymptotic charges, only one of the two solutions should be
stable. However, to see this physics, it may be necessary to include
the effects of the non-Abelian gauge fields, as in~\cite{wijnholt},
which we do not do.} 

We are going to focus on the linearised stability analysis, but we
will begin by discussing the thermodynamic aspects.\footnote{One
interesting suggestion in~\cite{freed:bh} was that in some cases,
black hole solutions should exist only for temperatures greater than a
critical value. We will see that for the non-extremal enhan\c con,
solutions with a regular event horizon exist for only for values of
the non-extremality parameter greater than a critical value---that is,
for sufficiently large energies. There also appears to be a maximum
temperature for these solutions, but no minimum.} In section
\ref{sec:thermo} we will compare the entropies of the two solutions,
and see that the horizon branch has larger entropy at large mass, as
we would expect. We can also calculate the specific heat for the
horizon branch; for the branch with a shell, the ambiguity in the
division of energy between the shell and the black hole prevent us
from obtaining a well-defined answer for the specific heat.

Our main focus is to look for dynamical instabilities that could take
us from one branch to the other. We particularly want to see whether
there is an instability at some value of the energy which could take
us from the branch with a shell to the horizon branch, which we think
should be the physical solution at large energies.

In section~\ref{sec:anal}, we set up an appropriate ansatz for the
perturbations. We consider only perturbations which are spherically
symmetric in the transverse space and translationally invariant along
the branes, as we are looking for a transition between two solutions
which preserve these symmetries. We consider the most general ansatz
consistent with the assumed symmetries. This ansatz is slightly more
general than the ansatz for perturbations of the extreme enhan\c con
considered in~\cite{Maeda:stab}; we find that our more general ansatz
is necessary to obtain non-trivial solutions of the full set of field
equations. We use the remaining diffeomorphism freedom to reduce the
linearised equations of motion to four second-order equations
for four functions characterising the perturbation. One of these
equations is decoupled from the others.

In section~\ref{psi3}, we consider the stability to this decoupled
mode. This equation is in fact identical to the free scalar wave
equation in this background. Since the mode is not coupled to the
shell, it satisfies simple continuous matching conditions there. We
reduce the equation to a one-dimensional bound state problem, and find
that the potential is negative in a region near the shell, so one
might expect that there could be bound states (and hence an
instability). Nevertheless, we present a general argument that there can
never be an instability associated with this mode. The idea is that
since the equation is just the free wave equation, a constant function
is a solution. This implies that the bound state problem has a
zero-energy solution with no nodes, and as a consequence, there can be
no bound states. 

In this paper, we will not consider the solution of the other three
coupled equations. The boundary conditions at the shell will be more
complicated for these modes, and we will need to solve the equations
numerically to determine if there is any instability. This analysis
will be continued in a forthcoming paper~\cite{toapp}.

\section{The enhan\c con solutions}
\label{sec:solns}

The original repulson geometry~\cite{Behrndt:enh,Kallosh:enh} is
constructed by wrapping $N$ D$(p+4)$--branes on a K3 manifold of
volume $V$. We will also consider including $M$ D$p$--branes parallel
to the noncompact directions of the D$(p+4)$--branes. This leaves an
unwrapped $(p+1)$--dimensional worldvolume in the six non--compact
dimensions. There are $5-p$ non--compact spatial dimensions transverse
to the branes. We will consider the case $p=2$, so we have coordinates
$r,\theta,\phi$ in the transverse directions, and $x^\mu$, $\mu=0,1,2$
in the noncompact directions along the branes. The Einstein frame
metric and fields are
\begin{eqnarray}
ds^2 &=& Z_2^{-5/8} Z_6^{-1/8} \eta_{\mu\nu} dx^\mu dx^\nu +
Z_2^{3/8} Z_6^{7/8}(dr^2 + r^2 d\Omega)  + V^{1/2} Z_2^{3/8}
Z_6^{-1/8} ds^2_{\rm K3} \ 
,\nonumber
\\
e^{2\Phi } &=& g_s^2 {Z_2}^{1/2}{ Z_6}^{-3/2}\ , \nonumber\\
C_{(3)} &=& ({Z_2} g_s)^{-1} dx^0 \wedge dx^1 \wedge dx^2\ , \nonumber\\
C_{(7)} &=& ({Z_6} g_s)^{-1} dx^0 \wedge dx^1 \wedge dx^2 \wedge
V\,\varepsilon_{\rm K3}
\ , \label{outside}
\end{eqnarray}
where the harmonic functions are 
\begin{eqnarray}
Z_6 &=& 1+{r_6\over r}\ ,\quad Z_2 =1 -{r_2\over r}\ ,
\label{harms}
\end{eqnarray}
the parameters are related by
\begin{eqnarray}
\quad r_6 = {g_sN\alpha'^{1/2}\over 2} \ ,\quad r_2={V_*\over V} {r_6}
\left( 1 - {M \over N} \right) \ ,
\label{arrs}
\end{eqnarray}
and $d\Omega$ denotes the metric on the unit two-sphere.  The running
K3 volume is
\begin{equation}
V(r)=V{Z_2(r)\over Z_6(r)}\ .
\end{equation}
$V = V_* = (2\pi \sqrt{\alpha'})^4$ at the enhan\c con radius, 
\begin{equation}
 r_{\rm e} = {2V_*\over {V} - V_* } r_6 \left( 1 - {M \over 2N}
 \right) \ .
\label{theradius}
\end{equation}
The repulson singularity would occur at $r= r_2 < r_{\rm e}$. 

The enhan\c con mechanism discovered in~\cite{JPP:enh} resolves this
repulson singularity. The essence of the mechanism is that the
singularity can never be formed. If one tries to assemble the repulson
from well-separated branes, the constituent branes will stop behaving
as pointlike objects and smear out into extended solitons at a certain
distance from the would-be singularity; the sphere at this radius is
called the enhan\c con sphere. This effect is due to the appearance of
additional light degrees of freedom, enhancing the gauge symmetry from
$U(1)$ to $SU(2)$, when the K3 volume is $V=V_* = (2\pi
\sqrt{\alpha'})^4$. The metric outside the enhan\c con sphere is still
the repulson geometry, but the sources are distributed over the
sphere, leaving flat space inside and removing the singularity. This
is the enhan\c con geometry.

Although this singularity resolution depends on stringy physics,
namely the appearance of additional light degrees of freedom which are
not contained in the original supergravity description, it was found
in~\cite{JMPR:enh} that the appearance of a shell at the enhan\c con
radius can be understood from a purely supergravity argument. If we
imagine distributing the sources on a spherically symmetric shell, so
that the exterior geometry is the repulson, while the spacetime inside
the shell is flat, then the energy density of the shell will be
positive only if the shell lies outside the enhan\c con radius. Thus,
the enhan\c con radius provides a minimum position for the shell.

Thus, the above geometry does not always apply for all $r$. For $M>N$
there is no repulson singularity, and we can assemble sources to form
the geometry in (\ref{outside}).\footnote{For $M>2N$, there is no
enhan\c con, and we can form the above geometry by bringing the branes
in individually from infinity. For $N<M<2N$, D6-branes on their own
smear out at $r=r_{\rm e}$. We can still form the geometry
(\ref{outside}) if we first form D2-D6 bound states, which can be
brought to the origin.} For $M<N$ however, some of the D6-branes cease
to be pointlike before we reach the singularity at $r = r_2$, and will
form an enhan\c con shell. This geometry then applies only outside the
shell.

We will mainly be interested in $M<N$. We assume that all $M$
D2-branes coincide at the origin, along with $N'$ D6-branes, where $N'
\leq M$. The remaining $N-N'$ D6-branes lie on an enhan\c con
shell. The geometry inside the shell is
\begin{equation}
g_s^{1/2}\,ds^2 = H_2^{-5/8} H_6^{-1/8} \eta_{\mu\nu} dx^\mu dx^\nu +
H_2^{3/8} H_6^{7/8} (dr^2 + r^2 d\Omega) + V^{1/2}
H_2^{3/8} H_6^{-1/8} ds^2_{K3}
\end{equation}
and the non--trivial fields are
\begin{eqnarray}
e^{2\Phi} &=& g_s^2 H_2^{1/2} H_6^{-3/2}\ , \nonumber\\
C_{(3)} &=& (g_s H_2)^{-1} dx^0
\wedge dx^1 \wedge dx^2\ ,\nonumber\\ 
C_{(7)} &=& (g_s H_6)^{-1} dx^0\wedge dx^1 \wedge dx^2
\wedge V\,\varepsilon_{\rm K3}\ ,
\end{eqnarray}
where
\begin{eqnarray}
H_2 = 1-{r_2 - r_2'\over r_{\rm e}} - {r_2' \over r}, 
\qquad r_2^\prime= r_6{V_*\over V}{N'-M\over N}\ ,
 \label{h2}\\
H_6 = 1+{r_6 - r_6'\over r_{\rm e}} + {r_6' \over r}, 
\qquad r_6^\prime= r_6{N'\over N}={g_sN'\alpha^{\prime1/2}\over2}\ .
 \label{h6}
\end{eqnarray}
The constant terms in the harmonic functions are chosen to ensure
continuity of the solution at the shell.

The supergravity argument can be extended to non-extremal
generalisations of the enhan\c con solution, which are difficult to
study from the string theory point of view. A non-extremal solution
was first written down in~\cite{JPP:enh}. In~\cite{JMPR:enh}, it was
found that there are two branches of non-extremal solutions, arising
from an ambiguity of a choice of sign in the solution of the
supergravity equations for the usual ansatz. The non-extremal
generalisation of the exterior geometry is
\begin{eqnarray}
g_s^{1/2}\,ds^2  &=&Z_2^{-5/8}{Z}_6^{-1/8} (-K dt^2 + dx_1^2+ dx_2^2) +
Z_2^{3/8}Z_6^{7/8} (K^{-1} dr^2 + r^2 d\Omega_2^2) \nonumber\\
&\phantom{+}&\hskip5cm + V^{1/2} Z_2^{3/8}Z_6^{-1/8}ds_{\rm K3}^2\ ,
\label{junterior}
\end{eqnarray}
the dilaton and R--R fields are 
\begin{eqnarray}
e^{2\Phi} &=& g_s^2Z_2^{1/2}Z_6^{-3/2}\ , \nonumber \\
C_{(3)} &=& (g_s \alpha_2 Z_2)^{-1}dt \wedge dx^1 \wedge dx^2\ ,
\nonumber\\
C_{(7)} &=& (g_s \alpha_6 Z_6)^{-1} dt \wedge dx^1 \wedge dx^2 \wedge
V\,\varepsilon_{\rm K3}\ , \label{fields}
\end{eqnarray}
and the various harmonic functions are given by
\begin{eqnarray}
K &=& 1 - {r_0 \over r}\ ,\nonumber\\ 
Z_2&=&1 + {\hat{r}_2 \over r}\ \qquad Z_6= 1 + {\hat{r}_6 \over r}\ .
\label{z6ext}
\end{eqnarray}
Here 
\begin{equation}
\hat{r}_6 = - {r_0 \over 2} + \sqrt{r_6^2 + \left( {r_0 \over 2} \right)^2}\ ,
\end{equation}
and $\alpha_6 = \hat{r}_6/r_6$. There are two choices for $\hat{r}_2$
consistent with the equations of motion:
\begin{equation}
\hat{r}_2 =  - {r_0 \over 2} \pm
\sqrt{r_2^2 + \left( {r_0 \over 2} \right)^2}\ ,
\label{choice}
\end{equation}
and $\alpha_2 = \hat{r}_2/r_2$. Here, $r_2$ and $r_6$ are still given
by (\ref{arrs}). We have changed our conventions for $Z_2$ to
facilitate comparison of formulae involving $Z_2$ and $Z_6$, so the
repulson singularity, if there is one, is at $r= - \hat{r}_2$.

There are two branches of solutions. For the upper sign in
(\ref{choice}), $\hat{r}_2 >0$, so there is no repulson singularity,
and the solution has a regular horizon at $r=r_0$. For the lower
choice of sign, however, the repulson singularity always lies outside
the would-be horizon, $|\hat{r}_2| > r_0$, and the geometry will be
corrected by an enhan\c con shell. We therefore refer to the former as
the `horizon branch' and the latter as the `shell branch'. It is
interesting that the appearance of a repulson singularity in the
non-extremal solutions is not connected to whether $M>N$, but rather
to a discrete choice. For $M>N$, the extremal solution is the same as
the solution at $r_0=0$ on the horizon branch, and we regard the
horizon branch as the only physical choice. For $M<N$, on the other
hand, the extremal solution is the solution at $r_0=0$ on the shell
branch, so we need to consider both branches of solutions. We will
henceforth focus on the case where $M<N$.

The shell branch exterior solution is cut off by an enhan\c con shell at
\begin{equation}
r_{\rm e} = {V_* \hat{r}_6 - V \hat{r}_2 \over V - V_*}\ .
\label{junktre}
\end{equation}
As in the extremal case, this shell will contain $N-N'$ D6-branes,
while the interior solution with $M$ D2-branes and $N'<M$ D6-branes is
\begin{eqnarray}
g_s^{1/2}\,ds^2 &=& H_2^{-5/8} H_6^{-1/8} \left(- {K(r_{\rm e})\over
L(r_{\rm e})}L dt^2 + dx_1^2+dx_2^2\right) + H_2^{3/8} H_6^{7/8}
(L^{-1} dr^2 + r^2 d\Omega) \nonumber\\
&\phantom{+}&\hskip5cm + V^{1/2} H_2^{3/8} H_6^{-1/8}ds^2_{K3}\ ,
\label{tag1}
\end{eqnarray}
with accompanying fields
\begin{eqnarray}
e^{2\Phi} &=& g_s^2 H_2^{1/2} H_6^{-3/2}\ , \nonumber\\
 \qquad C_{(3)} &=&
\left({K(r_{\rm e})\over L(r_{\rm e})}\right)^{1/2} (g_s
 \alpha'_2H_2)^{-1} dt \wedge dx^1 \wedge dx^2\ , \nonumber\\
C_{(7)} &=& \left({K(r_{\rm e})\over L(r_{\rm e})}\right)^{1/2}
(g_s \alpha'_6 H_6)^{-1} dt\wedge dx^1 \wedge dx^2 
\wedge V\,\varepsilon_{\rm K3}\ ,
\end{eqnarray}
where
\begin{eqnarray}
L &=& 1 - {r_0' \over r}\ ,\nonumber\\
H_2 &=& 1 +{\hat{r}_2-\hat{r}_2'\over r_{\rm e}} +
{\hat{r}_2' \over r}\ ,
\nonumber\\
&&\qquad\qquad\hat{r}_2' = - {r_0' \over 2} +\sqrt{r_2'^2 + 
\left( {r_0' \over 2} \right)^2}\ ,
\qquad r_2'=r_6{V_*\over V}{M-N'\over N}\ ,
\nonumber\\
H_6 &=& 1 +{\hat{r}_6-\hat{r}_6'\over r_{\rm e}} +
{\hat{r}_6' \over r}\ ,
\nonumber\\
&&\qquad\qquad\hat{r}_6' = - {r_0' \over 2} +\sqrt{r_6'^2 + 
\left( {r_0' \over 2} \right)^2}\ ,
\qquad r_6'=r_6{N'\over N}\ ,
\label{tag2}
\end{eqnarray}
and $\alpha_6' = \hat{r}_6'/r_6'$, $\alpha_2' = \hat{r}_2' / r_2'$. 
Note that we have introduced an independent non--extremality
scale $r_0'$ for the interior solution. Implicitly $r_0'<r_{\rm e}$
in order that the interior black hole actually fits inside the
shell. We have taken the horizon branch for the interior solution, as
$N' < M$. 

The shell branch solutions have an additional parameter, $r_0'$, which
is not determined by the asymptotic charges of the solution. It was
argued in~\cite{JMPR:enh} that this was simply a weakness of the
supergravity excision procedure, and that a better understanding of
the physics of the shell should fix this parameter. We will not
attempt to resolve this issue in this paper, but will simply consider
the stability of the shell branch solutions for arbitrary $r_0' < r_0,
r_{\rm e}$.  

\section{Thermodynamics}
\label{sec:thermo}

We would like to briefly compare the behaviours of the two
branches. The ADM energy density for these solutions is
\begin{equation}
E = {(2r_0 + \hat{r}_2 + \hat{r}_6) \over 4 G},
\end{equation}
where $G$ is Newton's constant. For the horizon branch, this gives 
\begin{equation}
E_{hb} = {1 \over 4G} \left( r_0 + \sqrt{{r_0^2 \over 4} + r_2^2} +
\sqrt{{r_0^2 \over 4} + r_6^2} \right),
\end{equation}
while for the shell branch,
\begin{equation}
E_{sb} = {1 \over 4G} \left( r_0 - \sqrt{{r_0^2 \over 4} + r_2^2} +
\sqrt{{r_0^2 \over 4} + r_6^2} \right).
\end{equation}
The difference between the $r_0=0$ solutions is $\Delta E =
|r_2|/2G$. For $M<N$, we need to add this much energy to the extremal
solution before we can get solutions on the horizon branch. 

The entropy and temperature on the horizon branch are easily obtained
from the metric (\ref{junterior}), giving
\begin{equation}
S_{hb} = {A \over 4G} = {\pi r_0 \over G} (r_0 + \hat{r}_2)^{3/8} (r_0
+ \hat{r}_6)^{7/8},
\end{equation}
\begin{equation}
T_{hb} = {1 \over 4\pi (r_0 + \hat{r}_2)^{1/2} (r_0 + \hat{r}_6)^{1/2}}. 
\end{equation}
For the shell branch, we must use the interior solution (\ref{tag1}),
which gives
\begin{equation}
S_{sb} = {A \over 4G} = {\pi {r_0'}^2 \over G} H_2(r_0')^{3/8}
H_6(r_0')^{7/8},
\end{equation}
\begin{equation}
T_{sb} = {1 \over 4\pi r_0'} \left( {K(r_{\rm e}) \over L(r_{\rm e})
H_2(r_0') H_6(r_0')} \right)^{1/2} . 
\end{equation}
On the horizon branch, we see that the temperature is a monotonic
function of $r_0$, and hence the specific heat is always
negative.\footnote{As a consequence, the conjecture
of~\cite{gubser:mitra} presumably implies that if the $x_1, x_2$
directions are non-compact, this solution has a Gregory-Laflamme type
instability~\cite{GL}. This is not the instability we are interested
in considering, as it seems unlikely to mediate a transition between
our two families of translationally-invariant solutions.} For the
shell branch, we cannot evaluate the specific heat, as we do not know
$r_0' (r_0)$.

The ambiguity in the interior solution on the shell branch prevents us
from comparing the entropies of the two solutions for most values of
the parameters. However, we can make a comparison at large energies,
when $r_0 \gg r_2, r_6$. Then 
\begin{equation}
E_{hb} \approx {r_0 \over 2G}, \quad S_{hb} \approx {\pi r_0^2 \over
G} \approx 4\pi G E_{hb}^2,
\end{equation}
as for an uncharged black hole, while for the shell branch,
\begin{equation}
E_{sb} \approx {r_0 \over 4G}, \quad S_{sb} \approx {\pi r_0'^2 \over
G} \left( {V_* \over V} \right)^{3/8} \approx 16 \pi G E_{sb}^2 {r_0'^2
\over r_0^2} \left( {V_* \over V} \right)^{3/8}.
\end{equation}
Since $r_0' < r_0$ and $V_*/V$ is a small parameter, we conclude that
the entropy is larger on the horizon branch at large mass. Thus, at
least for large fixed mass, we would expect the horizon branch to
dominate. 

It would also be interesting to compare the entropies at fixed low 
temperature (so again $r_0 \gg r_2, r_6$). Unfortunately, this is not
so straightforward. On the horizon branch, 
\begin{equation}
T_{hb} \approx {1 \over 4\pi r_0 }, \quad S_{hb} \approx {1 \over 16
\pi G T^2},
\end{equation}
but on the shell branch,
\begin{equation}
T_{sb} \approx {1 \over 4\pi r_0'}  \left( 1 - {r_0' \over r_{\rm e}}
\right)^{-1/2}, 
\end{equation}
so 
\begin{equation}
S_{sb} \approx {\pi r_0'^2 \over G} \left( 1 - {r_0 \over r_{\rm e}}
\right)^{3/8} \approx {1 \over 16 \pi G T^2} \left( 1 - {r_0 \over r_{\rm e}}
\right)^{3/8} \left( 1 - {r_0' \over r_{\rm e}} \right)^{-1}.
\end{equation}
Thus, whether $S_{sb}$ is smaller or larger than $S_{hb}$ in this
regime depends on how close $r_0'$ can be to $r_0$. Surprisingly, if
it is sufficiently close, $S_{sb}$ can be the larger. 

Thus, we see that thermodynamic considerations suggest that at
least for large masses, the horizon branch should be the preferred
one. Detailed investigation of the thermodynamics is hampered by the
fact that we don't know how $r_0'$ varies with $r_0$. Black hole
thermodynamics depends on studying the static vacuum solutions as
functions of the parameters, so the presence of an unphysical
one-parameter ambiguity in our family of solutions is a serious
impediment. 

\section{Perturbation ansatz}
\label{sec:anal}

We now turn to our main objective, the consideration of the stability of
these solutions. We wish to consider the simplest set of linearised
perturbations of the enhan\c con solutions which could provoke a
transition between the two branches. We will therefore assume that the
perturbations preserve many of the symmetries of the background
(\ref{junterior}). Specifically, we assume the spherical symmetry in
the $\theta, \phi$ directions, translational invariance in $x_1$ and
$x_2$, and the discrete symmetries under $x_1 \to -x_1$, $x_2 \to
-x_2$, $\phi \to -\phi$ are preserved. By a suitable choice of
coordinates, the most general perturbation consistent with these
symmetries can be written as the metric
\begin{eqnarray}
g_s^{1/2}\,ds^2  &=& e^{-\psi_1/2} \left[ \bar{Z}_2^{-1/2}
\bar{Z}_6^{-1/2} (-\bar{K} 
e^{\delta \psi_2} dt^2 + e^{-{1 \over 2} \delta \psi_2+ \delta \psi_3}
dx_1^2 + e^{-{1\over 2} \delta \psi_2 - \delta \psi_3} dx_2^2)
\right. \nonumber \\ &\phantom{+}& \left. +\bar{Z}_2^{1/2}
\bar{Z}_6^{1/2} (\bar{K}^{-1} dr^2 + r^2 d\Omega_2^2)  + V^{1/2}
\bar{Z}_2^{1/2} \bar{Z}_6^{-1/2}ds_{\rm K3}^2\right] \ ,
\label{pertext}
\end{eqnarray}
dilaton 
\begin{equation}
\bar{\phi} = \phi + \delta \phi, 
\end{equation}
and R--R fields 
\begin{equation}
\bar{C}_{(3)} = C_{(3)} + \delta C_{(3)}, \quad \bar{C}_{(7)} = C_{(7)} +
\delta{C}_{(7)}.   
\end{equation}
Here 
\begin{equation}
\psi_1 = \phi + \delta \psi_1,\  \bar{Z}_2 = Z_2 (1 + \delta Z_2),\ 
\bar{Z}_6 = Z_6 (1 + \delta Z_6),\  \bar{K} = K (1 + \delta K),
\end{equation}
the harmonic functions $Z_2, Z_6, K$ are as in (\ref{z6ext}), the
unperturbed dilaton $\phi$ is as in (\ref{fields}), and the R--R
potentials are as in (\ref{fields}). The first-order perturbations are
all functions of $(r,t)$ only, while the background quantities are
functions only of $r$.  We look for perturbations of the form $f(r)
e^{i \omega t}$.

Our ansatz is slightly more general than the ansatz adopted in the
study of perturbations of the extremal enhan\c con geometry
in~\cite{Maeda:stab}. We have introduced three new perturbation
functions, $\delta \psi_2$, $\delta \psi_3$, and $\delta K$. As we
will see shortly, we can choose to set $\delta K=0$ by a gauge
transformation. The first-order function $\delta \psi_3$ is the only
thing that breaks the rotational symmetry between $x_1$ and $x_2$. As
a consequence, it decouples from the other perturbations. We could set
it to zero without affecting the other modes; instead, we retain it,
and study it independently of the others. This provides us with a
single simple (but non-trivial) perturbation equation, which we study
in section \ref{psi3}. One might think that $\delta \psi_2$ would also
decouple, as it breaks the boost symmetry between $t$ and $x_1,x_2$
which (\ref{junterior}) respects. However, the assumption that the
perturbations depend on $t$ and not on $x_1,x_2$ also breaks this
symmetry, so we will find that $\delta \psi_2$ couples to the time
derivatives of the other perturbations, and it is not possible to set
it to zero. That is, it is necessary to consider the more general
ansatz containing $\delta \psi_2$ to satisfy all the field equations,
even in the extreme case.

We will now consider the full set of linearised equations for the
perturbations. The gauge field equations give 
\begin{equation} \label{gpert1}
\partial_r \delta C_{(3)} = -{1 \over 2} (4 \delta Z_2 + \delta \phi -
\delta \psi_1 ) \partial_r C_{(3)}
\end{equation}
and
\begin{equation} \label{gpert2}
\partial_r \delta C_{(7)} = -{1 \over 2} (4 \delta Z_6 - 3 \delta \phi
+ 3 \delta \psi_1) \partial_r C_{(7)}.
\end{equation}
The linear part of the stress tensor only involves $\partial_r \delta
C_{(3)}$ and $\partial_r \delta C_{(7)}$, so we can substitute
(\ref{gpert1},\ref{gpert2}) directly into the stress tensor. 

There are seven distinct equations coming from the linearised
Einstein's equations: six different diagonal components, and an
off-diagonal $[tr]$ component. With the dilaton equation, this gives
us eight equations; but with the gauge field perturbations fixed by
(\ref{gpert1},\ref{gpert2}), there are only seven undetermined
functions in our ansatz. The problem seems overdetermined, so it is
important to ask whether there will be any non-trivial solutions of
the full set of equations. We have written down the most general
perturbation consistent with the symmetries we have assumed, so we
expect there is sufficient redundancy in the equations to admit
non-trivial solutions.

In fact, we can see directly that there are non-trivial solutions to
these equations, using a trick from~\cite{HW}. We observe that the
ansatz (\ref{pertext}) does not completely fix the gauge, as there are
infinitesimal diffeomorphisms which preserve its form. Namely,
\begin{equation}
t \to t' = t + e^{i \omega t} \delta t(r), \quad r \to r' = r +
e^{i\omega t} \delta r(r),
\end{equation}
with 
\begin{equation}  
\partial_r \delta t = i\omega {Z_2 Z_6 \over K^2} \delta r . 
\end{equation}
If we apply this diffeomorphism to the non-extremal enhan\c con geometry
(\ref{junterior}), we obtain a metric of the form (\ref{pertext}) with
\begin{eqnarray} \label{partint}
\delta \psi_1^d &=& \left( \phi' - {4 \over 3r} \right) \delta r - {2
\over 3} \partial_r \delta r - {2 \over 3} i \omega \delta t, \\
\delta \psi_2^d &=& - {4 \over 3r} \delta r + {4 \over 3} \partial_r
\delta r + {4 \over 3} i \omega \delta t, \nonumber \\
\delta Z_6^d &=& \left( {Z_6' \over Z_6} + {2 \over r} \right) \delta r,
\nonumber \\
\delta Z_2^d &=& \left( {Z_2' \over Z_2} +{2 \over 3r} \right) \delta r
- {2 \over 3} \partial_r \delta r - {2 \over 3} i \omega \delta t,
\nonumber \\
\delta K^d &=& \left( {K' \over K} + {2 \over r} \right) \delta r - 2
\partial_r \delta r, \nonumber \\
\delta \phi^d &=& \phi' \delta r .\nonumber
\end{eqnarray}
Since this particular perturbation comes from a coordinate
transformation, it must solve the equations of motion. Thus, there are
non-trivial solutions of these equations. Of course, we are not
interested in solutions which are pure gauge, but this serves to
demonstrate that there is some redundancy in the equations. 

This diffeomorphism contains an arbitrary function; since we are not
interested in pure gauge perturbations, we should fix this additional
gauge symmetry. We can do so by setting one of the perturbations to
zero. It is convenient to choose $\delta K=0$. There remain
diffeomorphisms which will preserve $\delta K=0$: these have
\begin{equation} \label{rdif1}
\delta r = a r K^{1/2}
\end{equation}
and
\begin{eqnarray} \label{rdif2}
\delta t &=& i \omega a \left[-\frac{2(r_0+\hat{r}_2)(r_0+\hat{r}_6)}{\sqrt{K(r)}}+\left(\frac{r}{2}+\frac{7r_0}{4}+\hat{r}_2
+\hat{r}_6\right)r\sqrt{K(r)}+\right. \\ \nonumber
&&\left.+\left(\frac{15r_0^2}{8}+\frac{3r_0(\hat{r}_2+\hat{r}_6)}{2}+\hat{r}_2\hat{r}_6\right)
\ln{\frac{1+\sqrt{K(r)}}{1-\sqrt{K(r)}}} \right]+i\omega b
\end{eqnarray}

where $a$ and $b$ are arbitrary constants. The perturbations
(\ref{partint}) with this $\delta t$, $\delta r$ then give a
two-parameter family of solutions of the linearised equations with
$\delta K=0$. We will exploit this remaining coordinate freedom to
simplify the equations later.  

Having set $\delta K=0$, the contributions to the Ricci tensor linear
in the perturbations are
\begin{eqnarray}
\delta R_{tt} &=& {1 \over 4} (2\delta\ddot{\psi}_2 + 9 \delta\ddot{ \psi}_1
- 5\delta\ddot{ Z}_2 + 3\delta\ddot{ Z}_6) \\ \nonumber &&+ {K^2
\over 32 Z_2 Z_6} 
\left[ 16 \left(\delta \psi_2'' + {2 \over r} \delta \psi_2'\right) - 8 \left(\delta
\psi_1'' + {2 \over r} \delta \psi_1'\right) - 8 \left(\delta Z_2'' +
{2 \over r} \delta   Z_2'\right)  \right. \\ \nonumber &&-8
\left(\delta Z_6'' + 
{2 \over r} \delta Z_6'\right) +  16 \delta \psi_2' {K'
    \over K} + 4 \delta \psi_1' \left(-10 {K' \over K} + 5 {Z_2' \over Z_2}
  + {Z_6' \over Z_6}\right) \\ \nonumber &&
- \delta Z_2' \left( 5 {Z_2' \over Z_2} + {Z_6' \over     Z_6} \right) 
 + \delta Z_6' \left( -32 {K' \over K} + 15 {Z_2' \over Z_2} + 3
  {Z_6' \over Z_6} \right) \\ \nonumber && \left. + \left(\delta \psi_2 - \delta Z_2 -
\delta Z_6\right) \left( 10 {Z_2'^2 
    \over Z_2^2} - 10 {K' \over K} {Z_2' \over Z_2} +2 {Z_6'^2 \over
    Z_6^2} - 2 {K' \over K} {Z_6' \over Z_6}\right) \right] ,
\end{eqnarray}
\begin{eqnarray}
\delta_+ &=& \delta R_{tt} + {1\over 2} K \left(\delta R_{11} + \delta
R_{22}\right)  = {1 \over 4} \left(\delta\ddot{ \psi}_2 +8\delta\ddot{
\psi}_1 - 6\delta\ddot{   Z}_2 + 2\delta\ddot{ Z}_6 \right) 
\\ \nonumber &&+ {K^2 \over 32 Z_2 Z_6} \left[ 24
  \left(\delta \psi_2'' + {2 \over r}\delta  \psi_2' \right) + {K' \over K} (24
\delta \psi_2' - 32 \delta \psi_1' -24 \delta Z_6' + 8 \delta Z_2')
\right. \\ \nonumber
&&+ \left. \delta \psi_2 \left(15 {Z_2'^2 \over Z_2^2} -
  15 {K' \over K} {Z_2' \over Z_2} + 3 {Z_6'^2 \over Z_6^2} - 3 {K'
    \over K} {Z_6' \over Z_6} \right) \right],
\end{eqnarray}
\begin{eqnarray}
\delta_- &=& K\left(\delta R_{11} - \delta R_{22}\right)
  =\delta\ddot{ \psi}_3 + {K^2 
  \over 8 Z_2 Z_6} \left[ -8 \left(\delta \psi_3'' + {2 \over r}
  \delta \psi_3'\right) \right. \nonumber \\ &&-\left. 8
  \delta \psi_3' {K' \over K} + \delta \psi_3 \left( -5 {Z_2'^2 \over Z_2^2} +
  5 {K' \over K} {Z_2' \over Z_2} - {Z_6'^2 \over Z_6^2} + {K'
    \over K} {Z_6' \over Z_6} \right)\right] ,
\end{eqnarray}
\begin{eqnarray}
\delta R_{tr} &=& {1 \over 8} \left[ 4 \delta\dot{ \psi}_2' + 16 \delta\dot{
    \psi}_1' - 8 \delta\dot{ Z}_2' + 8 \delta\dot{ Z}_6' -2
    \delta\dot{ \psi}_2 {K' \over K} + \delta\dot{ \psi}_1 \left( -8 {K' \over K} +
    5 {Z_2' \over Z_2} + {Z_6' \over Z_6} \right) \right. \nonumber
\\ &&+ \left. \delta\dot{ Z}_2  \left( 4
    {K' \over K} - 5 {Z_2' \over Z_2} - {Z_6' \over Z_6} \right) +
    \delta\dot{ Z}_6 \left( -4 {K' \over K} +
    3 {Z_2' \over Z_2} - {Z_6' \over Z_6} \right) \right] ,
\end{eqnarray}
\begin{eqnarray}
\delta R_{rr} &=& {Z_2 Z_6 \over 4 K^2} \left(-\delta\ddot{ \psi}_1 
+\delta\ddot{ Z}_2 +\delta\ddot{ Z}_6\right) 
\\ \nonumber &&+ {1 \over 32} \left[ 72 \delta
  \psi_1'' - 24 \delta Z_2'' + 40 \delta Z_6'' - 16 \delta \psi_2' {K'
    \over K} + \delta \psi_1' \left( {16 \over r} + 40 {K' \over K} -12
      {Z_2' \over Z_2} - 28 {Z_6' \over Z_6}\right) \right.
\\ \nonumber &&+ \left. \delta Z_2' \left( -{16
	\over r} - 27 {Z_2' \over Z_2} + {Z_6' \over Z_6}\right) + \delta
      Z_6' \left( - {16 \over r} + 32 {K' \over K} - 15 {Z_2' \over Z_2}
      -35 {Z_6' \over Z_6}\right) \right] ,
\end{eqnarray}
\begin{eqnarray}
\delta R_{\theta\theta} &=& {r^2 Z_2 Z_6 \over 4 K} \left(-\delta\ddot{
\psi}_1 +\delta\ddot{ Z}_2 +\delta\ddot{ Z}_6\right) 
\\ \nonumber &&+ {r^2 K \over 32}
\left[ 8 \delta \psi_1'' -8 \delta Z_2'' -8 \delta Z_6'' + \delta
  \psi_1' \left( {80 \over r} + 8 {K' \over K}
+12 {Z_2' \over Z_2} + 28 {Z_6' \over Z_6}\right) \right. 
\\ \nonumber && + \left. \delta Z_2' \left( -{32 \over
r} -8 {K' \over K} - 3 {Z_2' \over Z_2} -7 {Z_6' \over Z_6}\right) + \delta
Z_6' \left( {32 \over r}  -8 {K' \over K} +9 {Z_2' \over Z_2} +21 {Z_6'
  \over Z_6}\right) \right] ,
\end{eqnarray}
and
\begin{eqnarray}
\delta R_{mn}[K3] &=& {\sqrt{V} Z_2 \over 4 K} \left(-\delta\ddot{ \psi}_1
+\delta\ddot{ Z}_2 -\delta\ddot{ Z}_6\right) 
\\ \nonumber &&+ {\sqrt{V} K \over 32 Z_6}
\left[ 8 \left(\delta \psi_1'' + {2 \over r} \delta \psi_1'\right) -8 \left(\delta
Z_2''+ {2 \over r} \delta Z_2'\right) +8 \left(\delta Z_6'' + {2 \over r} \delta
Z_6'\right) \right. 
\\ \nonumber &&+  \delta \psi_1' \left( 8 {K' \over K} +12 {Z_2' \over
Z_2} -4 {Z_6' 
\over Z_6}\right) + \delta Z_2' \left( -8 {K' \over K} - 3 {Z_2' \over
Z_2} + {Z_6' \over Z_6}\right) \\ \nonumber && + 
\delta Z_6' \left( 8 {K' \over K} +9 {Z_2' \over Z_2} -3
{Z_6' \over Z_6}\right) 
\\ \nonumber &&+ \left. \delta Z_6 \left(-6 {Z_2'^2 \over Z_2^2} + 6 {K' \over
  K} {Z_2' \over Z_2} + 2 {Z_6'^2 \over Z_6^2} - 2 {K' \over K} {Z_6'
  \over Z_6} \right) \right] ,
\end{eqnarray}
where we have introduced certain combinations which simplify the
resulting equations, $\dot{}$ signifies $\partial_t$, and $'$
signifies $\partial_r$. 

The linearized Einstein's equations give seven equations. First, there
are the simple equations from $\delta_-$ and $\delta_+$, which are
respectively
\begin{equation}
{Z_2 Z_6 \over K^2}\delta\ddot{ \psi}_3 - \delta \psi_3'' - \delta \psi_3' ({2 \over r} +
{K' \over K}) =0 \label{psi3eq}
\end{equation}
and
\begin{equation} \label{ee1}
{Z_2 Z_6 \over K^2} (\delta\ddot{ \psi}_2 + 8\delta\ddot{ \psi}_1 - 6
\delta\ddot{ Z}_2 + 2\delta\ddot{ Z}_6) + 3 \delta \psi_2'' + 3 \delta \psi_2' ({2
\over r} + {K' \over K}) + (-4 \delta \psi_1' -3 \delta Z_6' + \delta
Z_2') {K' \over K} =0.
\end{equation}
There are three more independent second-order equations, 
\begin{eqnarray} \label{ee2}
-2 {Z_2 Z_6 \over K^2}\delta\ddot{ Z}_6 + 2 \delta Z_6'' + \delta
 Z_6' \left(-{2 \over r} +2 {K' \over K} - 3{Z_6' \over Z_6}\right) +
 \delta \psi_1' \left(-{8 \over r} - 4{Z_6' \over Z_6}\right) &&
 \\\nonumber + \delta
 Z_2' \left({2 \over r} +{K' \over K} + {Z_6' \over Z_6}\right)
  + \delta Z_6 {3 \over 4} \left( {Z_6'^2 \over
 Z_6^2} - {K' \over K} {Z_6' \over Z_6} \right) + (3 \delta \phi +
 \delta \psi_1 - \delta Z_2) \left( {Z_6'^2 \over Z_6^2} - {K' \over
 K} {Z_6' \over Z_6} \right) &=& 0,
\end{eqnarray} 
\begin{eqnarray} \label{ee3}
-2 {Z_2 Z_6 \over K^2} (4\delta\ddot{ \psi}_1 + 2\delta\ddot{ \psi}_2 +
\delta\ddot{ Z}_6) + 6 (\delta Z_2'' + {2 \over r} \delta Z_2') +
 \delta \psi_1' \left(4 {K' \over K} -12 {Z_2' \over Z_2} \right) &&\\
 \nonumber +
 \delta Z_2' \left(5 {K' \over K} +3 {Z_2' \over Z_2} \right)+
 \delta Z_6' \left(3 {K' \over K} -9 {Z_2' \over Z_2}
 \right) -3 (\delta \phi - 5 \delta \psi_1 + 5 \delta Z_2 - 3 \delta
 Z_6) \left( {Z_2'^2 \over Z_2^2} - {K' \over K} {Z_2' \over Z_2}
 \right) &=& 0,
\end{eqnarray}
and
\begin{eqnarray} \label{ee4}
-8 {Z_2 Z_6 \over K^2}(7\delta\ddot{ \psi}_1 + 2\delta\ddot{ \psi}_2 - 3
\delta\ddot{ Z}_2 +\delta\ddot{ Z}_6) + 24 \delta \psi_1'' + \delta
 \psi_1' \left({144 \over r} + 40 {K' \over K} -12 {Z_2' \over Z_2}
 +36 {Z_6' \over Z_6}\right) && \\ + \delta Z_2' \left(-{24 \over r}
 -4 {K' \over K} + 3 {Z_2' \over Z_2} - 9 {Z_6' \over Z_6}\right) +
 \delta Z_6' \left({72 \over r} + 12 {K' \over K} -9 {Z_2' \over Z_2}
 +27 {Z_6' \over Z_6}\right) && \nonumber \\ +3 (-\delta \phi + 5
 \delta \psi_1 - 5\delta Z_2 + 3 \delta Z_6) \left( {Z_2'^2 \over
 Z_2^2} - {K' \over K} {Z_2' \over Z_2} \right) && \nonumber \\ \nonumber +9
 (-3\delta \phi - \delta \psi_1 + \delta Z_2 + \delta Z_6) \left(
 {Z_6'^2 \over Z_6^2} - {K' \over K} {Z_6' \over Z_6} \right) &=& 0.
\end{eqnarray}
The remaining Einstein's equations give us two equations which are
first-order in $\partial_r$. Integrating the $tr$ equation in $t$
gives
\begin{eqnarray} \label{fo1}
4\delta \psi_2' + 16 \delta \psi_1' - 8 \delta Z_2' + 8 \delta Z_6' - 2
\delta \psi_2 {K' \over K} + \delta \psi_1 \left(-8 {K' \over K} + 5{Z_2'
\over Z_2} + {Z_6' \over Z_6}\right)&&  \\ \nonumber + \delta Z_2
\left(4 {K' \over K} 
-5 {Z_2' \over Z_2} - {Z_6' \over Z_6}\right) + \delta Z_6 \left(-4
{K' \over K} + 3 {Z_2' \over Z_2} - {Z_6' \over Z_6}\right) + \delta
\phi \left(- {Z_2' \over Z_2} +3 {Z_6' \over Z_6}\right) &=& f(t),
\end{eqnarray}
and a suitable combination gives the equation
\begin{eqnarray} \label{fo2}
16 {Z_2 Z_6 \over K^2} \left( 4\delta\ddot{ \psi}_1 +\delta\ddot{ \psi}_2
-2\delta\ddot{ Z}_2 + 2\delta\ddot{ Z}_6 \right) -8 \delta \psi_2' {K' \over K} +
\delta \psi_1' \left(-{128 \over r} -32{K' \over K} -12 {Z_2' \over
Z_2} - 28 {Z_6' \over Z_6}\right) && \\ + \delta Z_2' \left({32
\over r} + 
16 {K' \over K} -12 {Z_2' \over Z_2} +4 {Z_6' \over Z_6}\right) +
\delta Z_6' \left(-{96 \over r} -16 {K' \over K} -12 {Z_2' \over Z_2}
- 28{Z_6' \over Z_6}\right)&&  \nonumber \\ + \delta \phi' \left(-4 {Z_2' \over Z_2}
+12 {Z_6' \over Z_6}\right) + (\delta \psi_1 -\delta Z_2) \left(
20{Z_2'^2 \over Z_2^2} - 20 {K' \over K} {Z_2' \over Z_2} + 4 {Z_6'^2
\over Z_6^2} - 4 {K' \over K} {Z_6' \over Z_6} \right) &&
\nonumber \\ \nonumber + \delta Z_6 \left(-3 {Z_2'^2 \over Z_2^2} +3  {K' \over
  K} {Z_2' \over Z_2} -4  {Z_6'^2 \over Z_6^2} +4 {K' \over K} {Z_6'
  \over Z_6} \right)  + \delta \phi \left( -4{Z_2'^2 \over Z_2^2} +4  {K' \over
  K} {Z_2' \over Z_2} +24  {Z_6'^2 \over Z_6^2} - 24 {K' \over K} {Z_6'
  \over Z_6} \right) &=& 0.
\end{eqnarray}

Finally, there is the dilaton equation 
\begin{eqnarray} \label{dileq}
-8{Z_2 Z_6 \over K^2}\delta\ddot{ \phi} + {8 \over r^2} \partial_r(K
 r^2 \partial_r \delta \phi) = (4 \delta \psi_1' - \delta Z_2' + 3
 \delta Z_6') \left( {Z_2' \over Z_2} -3 {Z_6' \over Z_6}\right)\\
 \nonumber + 3(3
 \delta \phi + \delta \psi_1 - \delta Z_6 - \delta Z_2) \left(
 {Z_6'^2 \over Z_6^2} -  {K' \over K} {Z_6'   \over Z_6} \right)  +
 (\delta \phi - 5 \delta \psi_1 + 5 \delta Z_2 - 3 \delta Z_6) 
 \left( {Z_2'^2 \over Z_2^2} -  {K' \over
  K} {Z_2' \over Z_2} \right) .
\end{eqnarray}

These equations are coupled in a complicated fashion, but we see that
as mentioned earlier, there is one simple equation, (\ref{psi3eq}). In
fact, this is the free wave equation.  We will discuss the analysis of
this decoupled mode in detail in section \ref{psi3}.

To simplify the other equations, we exploit the remaining
two-parameter family of diffeomorphisms
(\ref{rdif1},\ref{rdif2}). These can be used to construct a change of
variables which will simplify the equations: we replace $a$ and $b$
by functions $a(r)$ and $b(r)$, and set
\begin{eqnarray}
\delta \psi_1 &=& \delta \psi_1^d(a(r),b(r)), \\
\delta \psi_2 &=& \delta \psi_2^d(a(r),b(r)) + \Psi_2, \nonumber \\
\delta Z_6 &=& \delta Z_6^d(a(r),b(r)) + {\cal Z}_6, \nonumber \\
\delta Z_2 &=& \delta Z_2^d(a(r),b(r)), \nonumber \\
\delta \phi &=& \delta \phi^d(a(r),b(r)) + \Phi \nonumber,
\end{eqnarray}
with $\delta K=0$. The first term on the right-hand sides is the
diffeomorphism-induced perturbation (\ref{partint}) for the
diffeomorphism (\ref{rdif1},\ref{rdif2}), but with $a$ and $b$ now
functions. Since the diffeomorphism satisfies the equations of motion
for arbitrary constants $a$ and $b$, the linearised equations will
only involve derivatives of $a(r)$ and $b(r)$. The two first-order
equations (\ref{fo1},\ref{fo2}) can then be solved for $\partial_r
a(r)$ and $\partial_r b(r)$. Inserting these values into the other
four second-order equations (\ref{ee1}-\ref{ee4}) and the dilaton
equation (\ref{dileq}) gives two equations which are trivially
satisfied, and a coupled set of three second-order equations for
$\Psi_2$, ${\cal Z}_6$ and $\Phi$.  

It is convenient to write the coupled equations so that each one only
involves second derivatives of one of the functions. Then the equation
which involves $\Phi''$  is (where $'$ again denotes $\partial_r$, and we
assume that all the perturbations behave as $e^{i \omega t}$)
\begin{equation}
D (\Phi'' + {2r-r_0 \over r^2 K} \Phi' + {Z_2 Z_6 \over K^2} \omega^2
\Phi) + P^1_2 (\Psi_2' + 2 {\cal Z}_6') + Q^1_1 \Phi + Q^1_2 \Psi_2 +
Q^1_3 {\cal Z}_6 =0,
\end{equation}
with the polynomial coefficients
\begin{equation}
D = r^2 K (8 r^2 + 5
r \hat{r}_2  + 5 r \hat{r}_6 + 2 \hat{r}_2 \hat{r}_6 ) (4 r^2 + 3 r
\hat{r}_2 + 3 r\hat{r}_6  + 2 \hat{r}_2 \hat{r}_6),
\end{equation}
\begin{equation}
P^1_2 = -2 r^2 K ( - 2 r^2 \hat{r}_2 + 6r^2 \hat{r}_6 + 8 r \hat{r}_2 \hat{r}_6
+ 3 \hat{r}_2^2 \hat{r}_6  +  \hat{r}_2 \hat{r}_6^2 ),
\end{equation}
\begin{eqnarray}
Q^1_1 &=& -r^2 (4 r_0 \hat{r}_2 + 36 r_0 \hat{r}_6 + 3 \hat{r}_2^2 + 6
\hat{r}_2 \hat{r}_6 + 27 \hat{r}_6^2) \\ && \nonumber - r (40 r_0
\hat{r}_2 \hat{r}_6 
+ 2 \hat{r}_2^2 \hat{r}_6 +  30 \hat{r}_2 \hat{r}_6^2) 
-  12 r_0 \hat{r}_2^2 \hat{r}_6 - 8 \hat{r}_2^2 \hat{r}_6^2,
\end{eqnarray}
\begin{equation}
Q^1_2 = r_0 (-2 r^2 \hat{r}_2 + 6 r^2 \hat{r}_6 + 8 r \hat{r}_2
\hat{r}_6 + 3 \hat{r}_2^2 \hat{r}_6 + \hat{r}_2 \hat{r}_6^2), 
\end{equation}
\begin{eqnarray}
Q^1_3 &=& r^2 (8 r_0 \hat{r}_2 + 24 r_0 \hat{r}_6 + 9 \hat{r}_2^2 + 10
\hat{r}_2 \hat{r}_6 + 9 \hat{r}_6^2) \\ && \nonumber + r (24 r_0
\hat{r}_2 \hat{r}_6 + 
6 \hat{r}_2^2 \hat{r}_6 + 10 \hat{r}_2 \hat{r}_6^2)  + 6 r_0
\hat{r}_2^2 \hat{r}_6 -2 r_0 \hat{r}_2 \hat{r}_6^2.
\end{eqnarray}
The equation involving $\Psi_2''$ is
\begin{equation}
D (\Psi_2'' + {Z_2 Z_6 \over K^2} \omega^2 \Psi_2) + P^2_2 \Psi_2' + P^2_3
{\cal Z}_6' + Q^2_1 \Phi + Q^2_2 \Psi_2 + Q^2_3 {\cal Z}_6 = 0,
\end{equation}
where $D$ is as before, and the other polynomial  coefficients are
\begin{eqnarray}
P^2_2 &=& 64 r^5 + r^4(-32 r_0 + 120 \hat{r}_2 + 88\hat{r}_6) \\ &&
\nonumber + r^3 (-76
r_0  \hat{r}_2 -44 r_0 \hat{r}_6 +30 \hat{r}_2^2 + 172 \hat{r}_2
\hat{r}_6 + 30 \hat{r}_6^2) \\ && \nonumber + r^2 (-15 r_0 \hat{r}_2^2 -118 r_0
\hat{r}_2 \hat{r}_6 -15 r_0 \hat{r}_6^2 + 44 \hat{r}_2^2 \hat{r}_6 +
52 \hat{r}_2 \hat{r}_6^2) \\ && \nonumber + r (-28 r_0 \hat{r}_2^2 \hat{r}_6 -36 r_0
\hat{r}_2 \hat{r}_6^2 + 8 \hat{r}_2^2 \hat{r}_6^2) - 4 r_0 \hat{r}_2^2
\hat{r}_6^2 ,
\end{eqnarray}
\begin{equation}
P^2_3 = -8 r^2 \hat{r}_2 K (8r^2 + 16 r \hat{r}_6 + 3 \hat{r}_2 \hat{r}_6 +
5 \hat{r}_6^2), 
\end{equation}
\begin{equation}
Q^2_1 = 4 \hat{r}_2 (r^2 (-8 r_0 - 6 \hat{r}_2 -6 \hat{r}_6) + r (4
r_0 \hat{r}_6 - 7 \hat{r}_2 \hat{r}_6 + 3 \hat{r}_6^2) + 6 r_0
\hat{r}_2 \hat{r}_6 +2 \hat{r}_2\hat{r}_6^2),
\end{equation}
\begin{equation}
Q^2_2 = -2 r_0 \hat{r}_2 (8r^2 + 16 r \hat{r}_6 + 3 \hat{r}_2
\hat{r}_6 + 5 \hat{r}_6^2),
\end{equation}
\begin{equation}
Q^2_3 = 4 \hat{r}_2 (r^2(16 r_0 + 18 \hat{r}_2 + 2 \hat{r}_6) + r (12
r_0 \hat{r}_6  +21 \hat{r}_2 \hat{r}_6 - \hat{r}_6^2) -3 r_0 \hat{r}_2
\hat{r}_6 + 5 r_0 \hat{r}_6^2 + 6 \hat{r}_2 \hat{r}_6^2).
\end{equation}
The equation involving ${\cal Z}_6''$ is 
\begin{equation}
D ({\cal Z}_6'' + {Z_2 Z_6 \over K^2} \omega^2 {\cal Z}_6) + P^3_2
\Psi_2' + P^3_3 {\cal Z}_6' + Q^3_1 \Phi + Q^3_2 \Psi_2 + Q^3_3 {\cal
Z}_6 =0,
\end{equation}
where $D$ is as before, and the other polynomial  coefficients are
\begin{equation}
P^3_2 = -2r^2 K (6 r^2 \hat{r}_2 -2 r^2 \hat{r}_6 + 8 r \hat{r}_2
\hat{r}_6 + \hat{r}_2^2 \hat{r}_6 + 3 \hat{r}_2\hat{r}_6^2),
\end{equation}
\begin{eqnarray}
P^3_3 &=& 64 r^5 + r^4 (-32 r_0 + 64 \hat{r}_2 + 96 \hat{r}_6) \\ &&
\nonumber + r^3
(-20 r_0 \hat{r}_2 -52 r_0 \hat{r}_6 + 30 \hat{r}_2^2 +76 \hat{r}_2
\hat{r}_6 + 30 \hat{r}_6^2) \\ \nonumber &&+ r^2 (-15 r_0 \hat{r}_2^2 -22 r_0
\hat{r}_2 \hat{r}_6 -15 r_0 \hat{r}_6^2 + 28 \hat{r}_2^2 \hat{r}_6 +
20 \hat{r}_2 \hat{r}_6^2) \\ \nonumber && + r (-12 r_0 \hat{r}_2^2
\hat{r}_6 -4 r_0 
\hat{r}_2 \hat{r}_6^2 + 8 \hat{r}_2^2 \hat{r}_6^2) -4 r_0 \hat{r}_2^2
\hat{r}_6^2,
\end{eqnarray}
\begin{equation}
Q^3_1 = r^2(12 r_0 \hat{r}_2 + 12 r_0 \hat{r}_6 + 9 \hat{r}_2^2 + 10
\hat{r}_2 \hat{r}_6 + 9 \hat{r}_6^2) + r ( 8 r_0 \hat{r}_2 \hat{r}_6 +
10 \hat{r}_2^2 \hat{r}_6 + 6\hat{r}_2 \hat{r}_6^2) -4 r_0 \hat{r}_2^2
\hat{r}_6, 
\end{equation}
\begin{equation}
Q^3_2 = r_0 (6 r^2 \hat{r}_2 -2 r^2 \hat{r}_6 + 8 r \hat{r}_2
\hat{r}_6 + \hat{r}_2^2 \hat{r}_6 + 3 \hat{r}_2 \hat{r}_6^2),
\end{equation}
\begin{eqnarray}
Q^3_3 &=& -r^2 (24 r_0 \hat{r}_2 + 8 r_0 \hat{r}_6 + 27 \hat{r}_2^2 +
6 \hat{r}_2 \hat{r}_6 + 3 \hat{r}_6^2) \\ && \nonumber - r (24 r_0 \hat{r}_2 \hat{r}_6
+ 30 \hat{r}_2^2 \hat{r}_6 +2 \hat{r}_2 \hat{r}_6^2) + 2 r_0
\hat{r}_2^2 \hat{r}_6 - 6 r_0 \hat{r}_2 \hat{r}_6^2 - 8 \hat{r}_2^2
\hat{r}_6^2).  
\end{eqnarray}

Leaving aside the decoupled mode $\delta \psi_3$, which will be
discussed in the next section (and which we will find leads to no
instabilities), we have now reduced the perturbation problem to these
three second-order equations. The background whose stability we are
mainly interested in addressing is the shell branch solution, so we
will also need to formulate appropriate matching conditions at the
shell. The determination of the matching conditions and the numerical
investigation of the existence of suitable solutions of the equations
for negative $\omega^2$ will be the subject of a forthcoming companion
publication~\cite{toapp}.

\section{Stability of the free wave equation}
\label{psi3}

We will now discuss the stability of the solution against perturbation
by just turning on $\delta \psi_3$. We make the ansatz $\delta \psi_3
(t,r) = \Psi_3(r) e^{i \omega t}$. Then (\ref{psi3eq}) implies
\begin{equation}
{K^2 \over Z_2 Z_6} \left[ \partial_r^2 \Psi_3 + \left( {K' \over K}
+ {2 \over r} \right) \partial_r \Psi_3 \right] + \omega^2 \Psi_3
=0. \label{freeeq}
\end{equation}
If we consider the horizon branch, we need simply look for solutions
of this equation regular on the horizon and at infinity. For the shell
branch, (\ref{freeeq}) applies for $r > r_{\rm e}$, and 
\begin{equation}
{L^2 \over H_2 H_6} \left[ \partial_r^2 \Psi_3 + \left( {L' \over L}
+ {2 \over r} \right) \partial_r \Psi_3 \right] + {L(r_{\rm e}) \over K(r_{\rm
e})} \omega^2 \Psi_3 =0
\label{req}
\end{equation}
applies for $r< r_{\rm e}$. Since the shell does not couple to $\delta
\psi_3$, the appropriate boundary conditions at the shell are that
$\Psi_3$ and $\partial_r \Psi_3$ are continuous.

We translate this into a standard one-dimensional bound state problem,
by introducing the tortoise coordinate
\begin{equation}
r_* = \left\{ \begin{array}{c} \sqrt{  L(r_{\rm
e})\over K(r_{\rm e})} \int_{r_{\rm e}}^r {\sqrt{H_2 H_6} \over
L} d\bar{r} \quad r<r_{\rm e} , \\  \int_{r_{\rm e}}^r {\sqrt{Z_2 Z_6}
\over K} d\bar{r} \quad r > r_{\rm e} . \end{array} \right.  
\end{equation}
This coordinate runs from $-\infty$ at $r= r_0'$, to $+\infty$ at $r=
\infty$. We make a change of variable\footnote{Note that $\partial_r
\Psi_3$ being continuous is not equivalent to $\partial_r \psi$ being
continuous.}
\begin{equation}
\Psi_3 = \left\{ \begin{array}{c} {1 \over (Z_2 Z_6)^{1/4} r} \psi
\quad r<r_{\rm e}, \\
{1 \over (H_2 H_6)^{1/4} r} \psi  \quad r>r_{\rm e}. \end{array}
\right.
\end{equation}
Then (\ref{req}) becomes 
\begin{equation} \label{bst}
\partial_{r_*}^2 \psi + \omega^2 \psi + W \psi = 0,
\end{equation}
where for $r> r_{\rm e}$,
\begin{eqnarray}
W(r) = W_> &\equiv& {K^2 \over Z_2 Z_6} \left[ {1 \over 4} {Z_2'' \over
Z_2} + {1 \over 4} {Z_6'' \over Z_6} - {5 \over 16} \left( {Z_2' \over
Z_2} \right)^2 - {5 \over 16} \left( {Z_6' \over Z_6} \right)^2 - {1
\over 8} {Z_2' \over Z_2} {Z_6' \over Z_6} \right. \nonumber \\
 &&\left.+ {1 \over 4} \left( {Z_2'
\over Z_2} + {Z_6' \over Z_6} \right) {K' \over K} + {1 \over r} {K'
\over K} \right]\ ,
\label{vg}
\end{eqnarray}
while for $r< r_{\rm e}$,
\begin{eqnarray}
W(r) = W_< &\equiv& {K(r_{\rm e}) \over L(r_{\rm e})} {L^2 \over H_2
H_6} \left[ {1 \over 4} {H_2'' \over H_2} + {1 \over 4} {H_6'' \over
H_6} - {5 \over 16} \left( {H_2' \over H_2} \right)^2 - {5 \over 16}
\left( {H_6' \over H_6} \right)^2 - {1 \over 8} {H_2' \over H_2} {H_6'
\over H_6} \right. \nonumber \\
 &&\left.
+ {1 \over 4} \left( {H_2' \over H_2} + {H_6' \over H_6}
\right) {L' \over L} + {1 \over r} {L' \over L} \right]\ .
\label{vl}
\end{eqnarray}
Plugging in the functions from (\ref{z6ext}), we have 
\begin{eqnarray}
W_> &=& {K \over 16 Z_2^3 Z_6^3} \{ [8 (\hat{r}_2 + \hat{r}_6) + 16
r_0] r^{-3} + [3 \hat{r}_2^2 + 3 \hat{r}_6^2 + 30 \hat{r}_2 \hat{r}_6
+ 20 r_0 (\hat{r}_2+\hat{r}_6) ] r^{-4} \nonumber \\ &&+ [ 12
\hat{r}_2 \hat{r}_6 (\hat{r}_2 + \hat{r}_6) + 9 r_0 (\hat{r}_2+
\hat{r}_6)^2] r^{-5} \nonumber+ [ 4 \hat{r}_2^2 \hat{r}_6^2 + 8
r_0 \hat{r}_2 \hat{r}_6 (\hat{r}_2 \hat{r}_6)] r^{-6}  \\ &&+ 4
r_0 \hat{r}_2^2 \hat{r}_6^2 r^{-7} \}.
\end{eqnarray}
The general form of $W_<$ is complicated, but in the case $M=0$, where
we have simply an uncharged black hole inside the shell, 
\begin{equation}
W_< = {K(r_{\rm e}) \over Z_2(r_{\rm e}) Z_6(r_{\rm e})} {L \over
L(r_{\rm e})} {r_0' \over r^3}\ .
\end{equation}
On the horizon branch, where $\hat{r}_2 > 0$, $W >0$ everywhere, and
there can be no instability associated with this mode. This is as we
would expect; the horizon branch looks like a normal charged black
hole solution, and the free wave equation has no non-constant
solutions regular both on the horizon and at infinity. However, on the
shell branch, there may be a region with $W_> <0$. (Since we take
the horizon branch for the solution inside the shell, $W_<$ is always
positive.) The leading term is always positive, as
\begin{equation}
\hat{r}_2 + \hat{r}_6 + r_0 = {1 \over 2} \sqrt{ 4r_6^2 + r_0^2} \pm
{1 \over 2} \sqrt{ 4 r_2^2 + r_0^2} >0 ,
\end{equation}
since $|r_2| < r_6$. On the other hand, $W_>$ is always negative
near $r= -\hat{r}_2$. As $r \to - \hat{r}_2$,
\begin{equation}
W_> \approx - {5 \over 16} {\hat{r}_2^2 K^2 \over r^4 Z_2^3 Z_6} <0.
\end{equation} 
If we considered just the pure repulson solution, this divergence
would lead us to suspect the solution is unstable to a perturbation by
$\delta \psi_3$. Although one would need to consider the issue
of boundary conditions at the singularity, $W_>$ diverges sufficiently
quickly that there could be bound states supported away from
$r=- \hat{r}_2$. The question, then, is whether the enhan\c con excises
this instability, along with the various other undesirable features of
the geometry.

\begin{figure}[t]   
\begin{center}  
\includegraphics[width=0.4\textwidth,height=0.3\textheight]{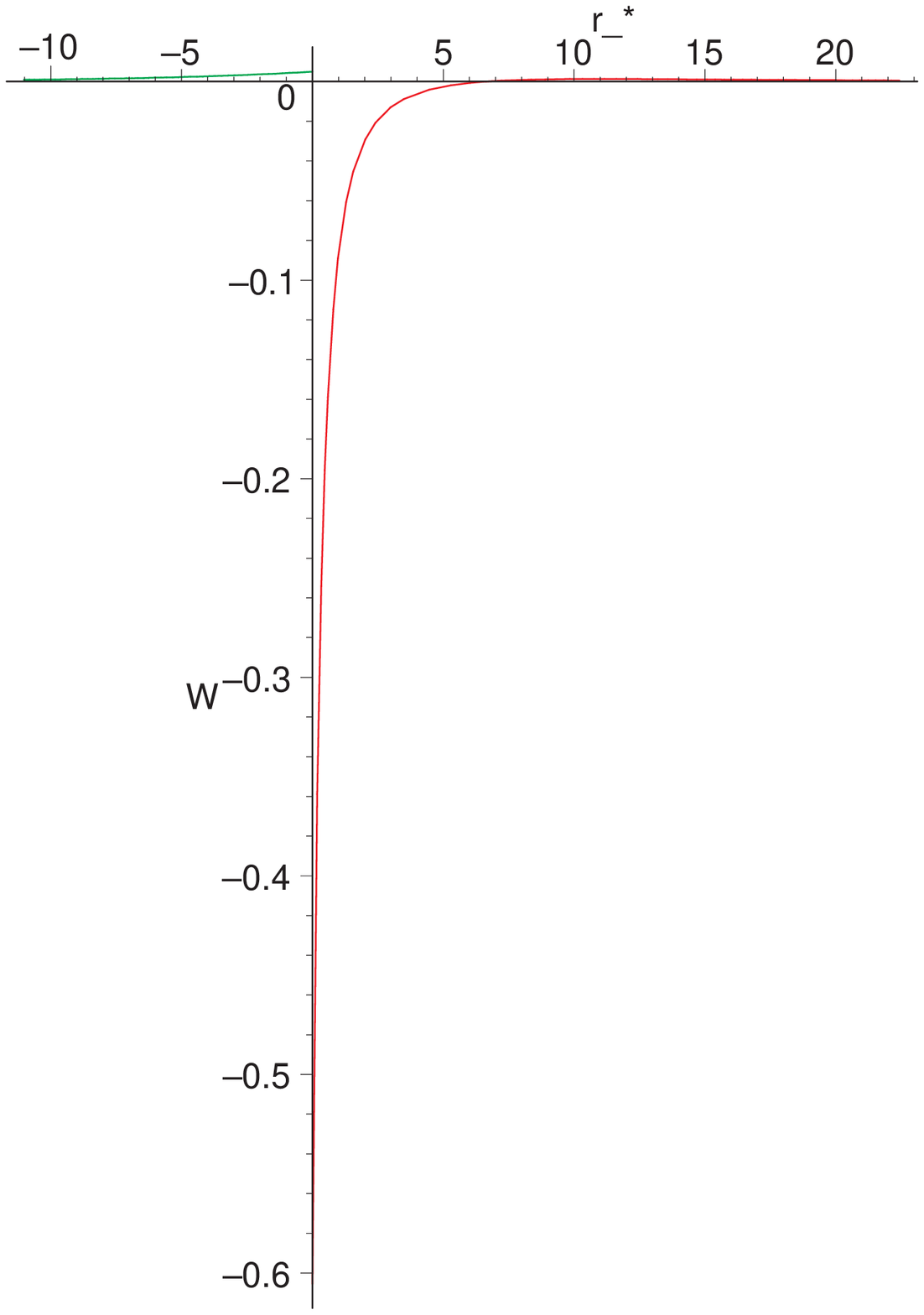}
\includegraphics[width=0.4\textwidth,height=0.3\textheight]{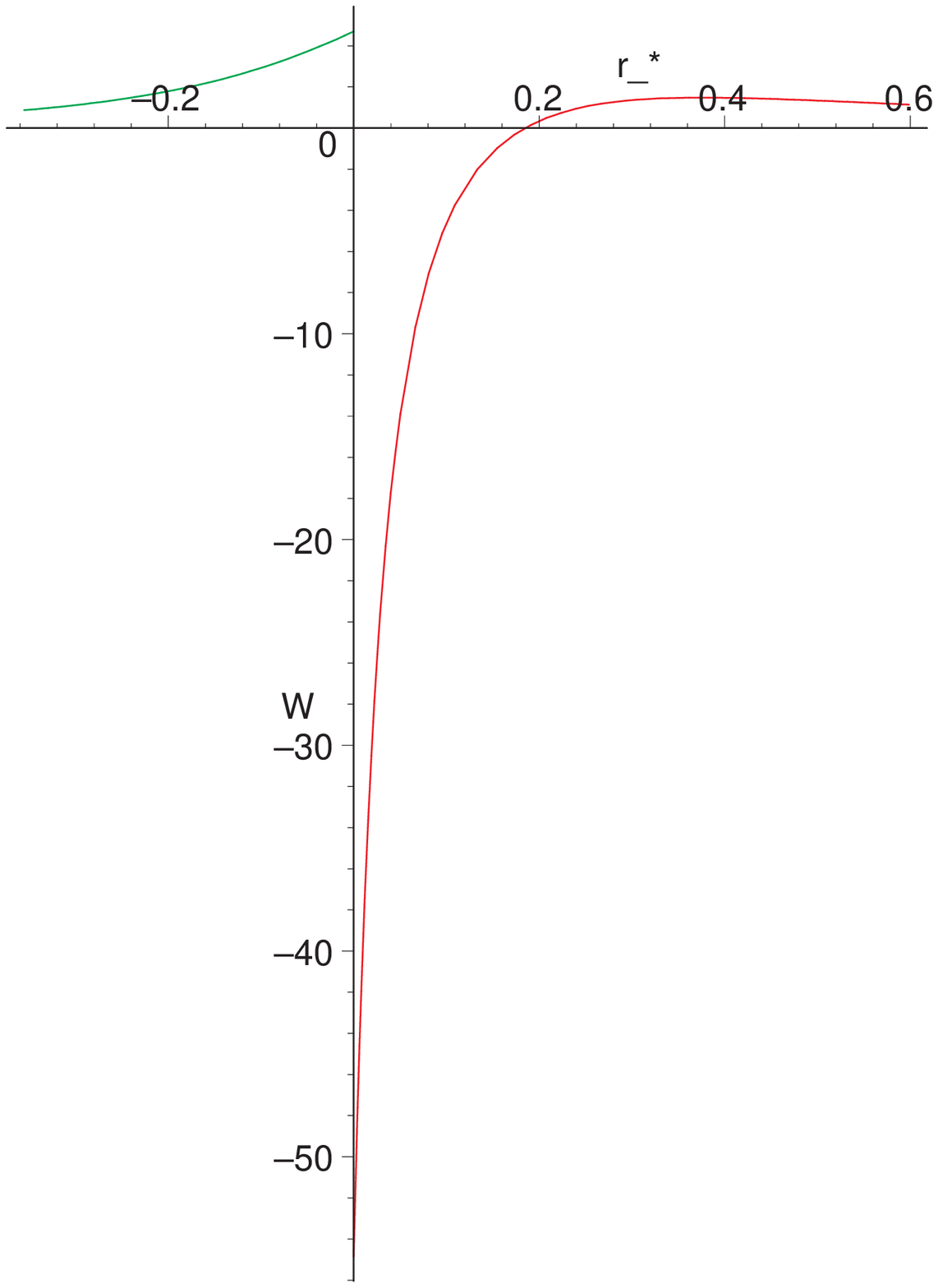}
\end{center}  
\caption{$ r_6^2 W$ plotted against $r_*/r_6$ for (left) $r_0 = 10 r_6$,
$V = 1000 V_*$, $M=0$, (right) $r_0 =
r_6/10$, $V = 1000 V_*$, $M=0$.   
} \label{fig:1}
\end{figure}

In figure \ref{fig:1}, we plot the potential for some
representative values of the parameters. We see that there is a
substantial region outside the shell where the potential is negative,
and might suspect that this signals an instability. 

However, there is a general argument which says that there can never
be an instability in this case~\cite{boundst}. First, we note that as
(\ref{freeeq}) is simply the free wave equation in this background, it
always has $\delta \psi_3 =$ constant as a solution. In terms of the
bound state problem (\ref{bst}), this translates into the statement
that there is a zero energy $(\omega=0)$ eigenmode $\psi_0$ of
(\ref{bst}) which of the same sign and is bounded everywhere; we can
take it to be always positive. This zero mode $\psi_0$ does not vanish
at the boundaries, so it is not a physical perturbation but it is
still an acceptable mathematical solution of this equation.

Now assume there is a discrete spectrum of bound states
$\psi_{\omega}$ with negative energy. These are our hypothetical
unstable modes with $\omega^2<0$.  We can see from the form of
(\ref{bst}) that they go to zero as $r_* \to \pm \infty$. This means
that they are bounded solutions and physical perturbations of our
problem. The standard `node rule' for the number of nodes of the
eigenfunctions of the discrete bound states says that in order of
increasing energy, the $n$th eigenmode has $n-1$ nodes (without
including the boundary ones).  Thus, the lowest negative mode
$\psi_{\omega_{max}}$ must have no nodes in the sense of the above
rule: we can take it to be everywhere positive.

Both $\psi_0$ and $\psi_{\omega_{max}}$ are solutions of the wave
equation (\ref{bst}). By multiplying the equation for each mode by the
other and taking the difference, and integrating over $r_*$, we can
obtain the equation
\begin{equation}
(\psi_{\omega_{max}}\partial_{r_{*}}\psi_0-\psi_0\partial_{r_{*}}\psi_{\omega_{max}})|_{r_*
    = \pm \infty}
    = - \omega_{max}^2 \int^{\infty}_{-\infty}\psi_{\omega_{max}}\psi_0dr_{*} 
\end{equation}
The left-hand side is the difference of the Wronskians calculated at
the boundaries. Since the eigenmode $\psi_0$ approaches a positive
constant at the boundaries $r_{*}=\pm\infty$, while the eigenmode
$\psi_{\omega_{max}}$ goes to zero, the Wronskian vanishes at each
boundary. Hence, the left-hand side is zero.  On the other hand, since
both $\psi_{\omega_{max}}$ and $\psi_0$ are supposed to be everywhere
positive, the right-hand side cannot be zero unless
$\omega_{max}=0$. Thus, assuming the existence of eigenmodes
$\psi_{\omega}$ with $\omega^2<0$ produces a contradiction. Hence
there can be no such modes, implying that the geometry is stable to 
perturbation by $\delta \psi_3$.

\centerline{\bf Acknowledgements}
\medskip    
    
We are grateful to James Gregory, Clifford Johnson
and Matt Strassler for useful discussions and to  Geoffrey Potvin for  bringing to our notice a typo in (\ref{rdif2}).
AD is supported in part by EPSRC studentship
00800708 and by a studentship from the University of Durham. SFR is
supported by an EPSRC Advanced Fellowship.

\bibliographystyle{/home/aplm/dma0sfr/tex_stuff/bibs/utphys}  
   
\bibliography{apostolos}   
    
\end{document}